\documentclass{article}
\usepackage{spconfa4,amsmath,graphicx}
\usepackage{cite}
\usepackage{siunitx}

\usepackage{pgfplots}
\usetikzlibrary{shapes, arrows, chains, calc}
\pgfplotsset{compat=newest, grid style={dotted}}
\graphicspath{{./img}}

\usepackage[nolist]{acronym}
\begin{acronym}
	\acro{stft}[STFT]{short-time Fourier transform}
	\acro{rtf}[RTF]{relative transfer function}
	\acro{atf}[ATF]{acoustic transfer function}
	\acro{snr}[SNR]{signal-to-noise ratio}
	\acro{psd}[PSD]{power spectral density}
	\acro{cs}[CS]{covariance subtraction}
	\acro{cw}[CW]{covariance whitening}
	\acro{mvdr}[MVDR]{minimum variance distortionless response}
	\acro{gevd}[GEVD]{generalised eigenvalue decomposition}
	\acro{evd}[EVD]{eigenvalue decomposition}
	\acro{pm}[PM]{power method}
	\acro{vad}[VAD]{voice activity detector}
	\acro{mwf}[MWF]{multichannel Wiener filter}
	\acro{doa}[DOA]{direction-of-arrival}
	\acro{sc}[SC]{spatial coherence}
	\acro{asn}[ASN]{acoustic sensor network}
\end{acronym}

\usepackage{amssymb}
\usepackage{empheq}
\usepackage{bm}
\usepackage{algorithmic}
\usepackage{algorithm}

\newcommand{\x}[0]{\mathbf{x}}
\newcommand{\y}[0]{\mathbf{y}}
\newcommand{\n}[0]{\mathbf{n}}
\newcommand{\h}[0]{\mathbf{h}}
\newcommand{\w}[0]{\mathbf{w}}
\newcommand{\bx}[0]{\bar{\mathbf{x}}}
\newcommand{\by}[0]{\bar{\mathbf{y}}}
\newcommand{\bn}[0]{\bar{\mathbf{n}}}
\newcommand{\bh}[0]{\bar{\mathbf{h}}}

\newcommand{\Rn}[0]{\mathbf{R}_\mathrm{n}}
\newcommand{\Rx}[0]{\mathbf{R}_\mathrm{x}}
\newcommand{\Ry}[0]{\mathbf{R}_\mathrm{y}}
\newcommand{\bRn}[0]{\bar{\mathbf{R}}_\mathrm{n}}
\newcommand{\bRx}[0]{\bar{\mathbf{R}}_\mathrm{x}}
\newcommand{\bRy}[0]{\bar{\mathbf{R}}_\mathrm{y}}
\newcommand{\hRn}[0]{\hat{\mathbf{R}}_\mathrm{n}}
\newcommand{\hRx}[0]{\hat{\mathbf{R}}_\mathrm{x}}
\newcommand{\hRy}[0]{\hat{\mathbf{R}}_\mathrm{y}}

\newcommand{\phixone}[0]{\Phi_\mathrm{x_1}}
\newcommand{\YE}[0]{Y_{\mathrm{E}}}
\newcommand{\XE}[0]{X_{\mathrm{E}}}
\newcommand{\NE}[0]{N_{\mathrm{E}}}

\newcommand{\eo}[0]{\mathbf{e}}
\newcommand{\beE}[0]{\bar{\mathbf{e}}_\mathrm{E}}
\newcommand{\beo}[0]{\bar{\mathbf{e}}}
\newcommand{\hcs}[0]{\mathbf{h}_\mathrm{CS}}
\newcommand{\hro}[0]{\mathbf{h}_\mathrm{R1}}
\newcommand{\hcw}[0]{\mathbf{h}_\mathrm{CW}}
\newcommand{\hpmcw}[0]{\mathbf{h}_\mathrm{PM-CW}}
\newcommand{\hpmcs}[0]{\mathbf{h}_\mathrm{PM-CS}}

\newcommand{\href}[0]{\mathbf{h}_\mathrm{REF}}
\newcommand{\hh}[0]{\hat{\mathbf{h}}}
 
 \usepackage[subtle]{savetrees}
\title{Relative Transfer Function Estimation Exploiting Spatially\\Separated Microphones in a Diffuse Noise Field}
\name{Nico G\"o\ss ling, Simon Doclo\thanks{This work was supported by the Collaborative Research Centre 1330 Hearing Acoustics, the Cluster of Excellence 1077 Hearing4all, funded by the German Research Foundation (DFG), and by the joint Lower Saxony-Israeli Project ATHENA.}}%
\address{University of Oldenburg, Department of Medical Physics and Acoustics and Cluster of Excellence\\
Hearing4All, Oldenburg, Germany}
\begin{document}
\ninept
\sloppy
\setlength{\belowdisplayskip}{6.5pt}
\setlength{\abovedisplayskip}{6pt}
\maketitle
\vspace{-0.2cm}
\begin{abstract}
Many multi-microphone speech enhancement algorithms require the relative
transfer function (RTF) vector of the desired speech source, relating
the acoustic transfer functions of all array microphones to a reference
microphone. In this paper, we propose a computationally efficient method
to estimate the RTF vector in a diffuse noise field, which requires
an additional microphone that is spatially separated from the microphone
array, such that the spatial coherence between the noise components in
the microphone array signals and the additional microphone signal is
low. Assuming this spatial coherence to be zero, we show that an
unbiased estimate of the RTF vector can be obtained. Based on real-world
recordings experimental results show that the proposed RTF estimator
outperforms state-of-the-art estimators using only the microphone array
signals in terms of estimation accuracy and noise reduction performance.
\end{abstract}
\vspace{-0.2cm}
\section{Introduction}
\label{sec:intro}
In many hands-free speech communication systems such as hearing aids, hearables or other assistive listening devices, the captured speech signal is often corrupted by additive background noise, such that speech enhancement methods are required to improve speech quality and speech intelligibillity \cite{Doclo2015}.
When more than one microphone is available, it is not only possible to exploit the spectro-temporal properties but also the spatial properties of the sound field to extract the desired speech source at a certain position from the noisy microphone signals.
By using spatially distributed microphones, e.g., one or more external microphones in addition to the microphones on the hearing device, the spatial sampling of the sound field can be increased \cite{Bertrand2009,Markovich-Golan2015,Szurley2016,Yee2017J,Goessling2017,Ali2018}.\\
A well-known multi-microphone speech enhancement method is the \ac{mvdr} beamformer \cite{Doclo2015,Veen1988}.
In a reverberant environment the MVDR beamformer either requires the \acp{atf} between the desired speech source and the microphones, which are difficult to accurately estimate in practice, or the \acp{rtf} of the desired speech source, which relate the \acp{atf} to a reference microphone \cite{Doclo2015,Gannot2001}.
Since \acp{rtf} can be exploited in many multi-microphone speech enhancement methods \cite{Warsitz2007,Markovich2009,Krueger2011,Marquardt2015a,Hadad2016,Hassani2016,Taseska2016,Koutrouvelis2018}, accurately estimating the \acp{rtf} of one or more sources is an important task.
In the literature several methods for estimating the \acp{rtf} have been proposed \cite{Gannot2001,Cohen2004,Warsitz2007,Markovich2009,Krueger2011,Markovich2015,Giri2016,Varzandeh2017}, where most recent methods are based either on \ac{cs} or \ac{cw}.
These methods usually require an estimate of the microphone signal covariance matrix (e.g., estimated during speech-plus-noise periods) and the noise covariance matrix (e.g., estimated during noise-only periods).
Although an iterative version of the \ac{cs} and \ac{cw} methods has been presented in \cite{Krueger2011,Varzandeh2017}, the computational complexity of the \ac{cs}-based and \ac{cw}-based \ac{rtf} estimation methods is generally high due to the involved matrix operations (possibly involving an \ac{evd}), which is especially relevant for an online-implementation.\\
In this paper, we propose a computationally efficient method to estimate the \acp{rtf} of a \textit{local} microphone array (e.g., on a hearing device) by exploiting the availability of an \textit{external} microphone that is spatially separated from the local microphone array.
We consider a diffuse noise field and assume that the distance between the external microphone and the local microphone array is large enough such that the \ac{sc} between the noise components in the local microphone signals and the external microphone signal is low.
When assuming this \ac{sc} to be zero, we show that a simple \ac{rtf} estimator can be derived that only depends on the microphone signal covariance matrix.
Based on real-world recordings with (local) head-mounted microphones and an (external) table microphone, we compare the performance of the proposed \ac{rtf} estimator and different CS-based and CW-based RTF estimators (using only the local microphone signals).
Simulation results show that the proposed estimator yields the best performance when used in an online implementation of the binaural MVDR beamformer.
%
%\vspace{-0.2cm}
\section{Signal model}
We consider an acoustic scenario with one desired speech source and diffuse noise (e.g., babble noise) in a reverberant environment.
The $m$-th microphone signal $Y_m(k,l)$ of an $M$-element \textit{local} microphone array can be written in the \ac{stft} domain as
\begin{equation}
	\label{eq:microphoneSignal}
	Y_m(k,l) = X_m(k,l) + N_m(k,l), \quad m \in \{1, \dots, M\},
\end{equation}
where $X_m(k,l)$ denotes the speech component, $N_m(k,l)$ denotes the noise component, and $k$ and $l$ denote the frequency and frame indices, respectively. For the sake of brevity, we will omit these indices in the remainder of the paper wherever possible.
All microphone signals can be stacked in a vector, i.e.,
\begin{equation}
\label{eq:y}
	\y = \left[ Y_1, \; Y_2, \; \dots, Y_M\right]^T \in \mathbb{C}^M,
\end{equation}
which can be written as
\begin{equation}
	\label{eq:vectorNotation}
	\y = \x + \n,
\end{equation}
where the speech vector $\x$ and the noise vector $\n$ are defined similarly as $\y$ in \eqref{eq:y}.
Without loss of generality, we choose the first microphone as the reference microphone.
The \ac{rtf} vector of the desired speech source is then given by
\begin{equation}
	\label{eq:rtf}
	\h = \left[1, \; \frac{A_2}{A_1}, \; \dots, \frac{A_M}{A_1}\right]^T,
\end{equation}
where $A_m$ is the \ac{atf} between the desired speech source and the $m$-th microphone. Using \eqref{eq:rtf}, the speech vector can be written as
\begin{equation}
	\label{eq:speechVector}
	\x = X_1\h.
\end{equation}
The speech covariance matrix $\Rx \in \mathbb{C}^{M\times M}$ and the noise covariance matrix $\Rn \in \mathbb{C}^{M \times M}$ are given by
\begin{align}
	\label{eq:corrMatRx}
	\Rx &= \mathcal{E}\{\x\x^H\} = \phixone \h\h^H,\\
	\Rn &= \mathcal{E}\{\n\n^H\},
	\label{eq:corrMatRn}
\end{align}
where $(\cdot)^H$ denotes complex conjugation, $\mathcal{E}\{\cdot\}$ denotes the expectation operator and $\phixone$ is the speech \ac{psd} in the first microphone.
Assuming statistical independence between the speech and the noise components, the covariance matrix of the microphone signals $\Ry = \mathcal{E}\{\y\y^H\}$ is equal to
\begin{equation}
	\label{eq:corrMatRy}
	\Ry = \Rx + \Rn.
\end{equation}
When applying a filter-and-sum beamformer $\w \in \mathbb{C}^M$ to the microphone signals, the output signal $Z$ is given by
\begin{equation}
	\label{eq:output}
	Z = \w^H\y.
\end{equation}
The \ac{mvdr} beamformer \cite{Doclo2015,Gannot2001} aims at minimizing the output noise \ac{psd} while preserving the speech component in the reference microphone signal and is hence given by
\begin{equation}
	\label{eq:mvdr}
	\begin{split}
		\w &= \arg \; \min_{\w} \; \w^H\Rn\w \quad \text{subject to} \quad \w^H\h = 1,\\
		&= \frac{\Rn^{-1}\h}{\h^H\Rn^{-1}\h}.
	\end{split}
\end{equation}
From \eqref{eq:mvdr} it is clear that the \ac{mvdr} beamformer only requires knowledge about the noise covariance matrix $\Rn$ and the \ac{rtf} vector of the desired speech source $\h$.
\vspace{-0.2cm}
\section{RTF vector estimation}
\label{sec:esti}
In this section, we briefly review two commonly used methods to estimate the \ac{rtf} vector $\h$, namely the \ac{cs} method\cite{Cohen2004,Serizel2014,Markovich2015} and the \ac{cw} method \cite{Serizel2014,Markovich2009}.
For both methods, we also discuss iterative versions based on the power iteration method \cite{Warsitz2007,Krueger2011,Varzandeh2017}.
All methods require an estimate of the microphone signal covariance matrix $\Ry$ and the noise covariance matrix $\Rn$, where $\hRy$ is estimated during speech-plus-noise frames and $\hRn$ is estimated during noise-only frames, assuming a \ac{vad} is available.
The computational complexity of the CS-based and CW-based methods depends on the required matrix operations, where especially matrix inversion or \ac{evd} will result in a large computational complexity.
\vspace{-0.2cm}
\subsection{Covariance subtraction (CS)}
\label{sec:cs}
By using the rank-1 model in \eqref{eq:corrMatRx}, the \ac{rtf} vector $\h$ can be calculated as any column of the speech correlation matrix $\Rx$, normalised by the entry corresponding to the reference microphone, i.e.,
\begin{equation}
	\label{eq:cs}
	\h_\mathrm{CS} = \frac{\Rx\eo}{\eo^T\Rx\eo},
\end{equation}
where $\eo = [1, \: 0, \: \dots, \: 0]^T$ is a selection vector consisting of zeros with one element equal to 1.
Usually, the speech covariance matrix $\Rx$ is estimated as $\hRx = \hRy - \hRn$.\\
Although the CS method has a relatively low computational complexity, its performance is not always very good since due to estimation errors the estimated speech covariance matrix $\hRx$ typically does not have rank-1 \cite{Serizel2014,Markovich2015}.
Hence, the RTF vector can also be estimated as the principal eigenvector (corresponding to the largest eigenvalue) of $\hRx$ normalising by the entry corresponding to the reference microphone. We denote this estimate as $\h_\mathrm{R1}$. It has been shown in \cite{Serizel2014} that $\hro$ outperforms $\hcs$ when used in a \ac{mwf}, but obviously has a larger computational complexity due to the \ac{evd}.
\vspace{-0.2cm}
\subsection{Covariance whitening (CW)}
By using a square-root decomposition (e.g., Cholesky decomposition) of the noise covariance matrix $\Rn$, i.e.,
\begin{equation}
	\label{eq:cholesky}
	\Rn = \Rn^{H/2}\Rn^{1/2},
\end{equation}
the pre-whitened microphone signal covariance matrix is given by
\begin{equation}
	\label{eq:prewhitening}
	\Ry^{\mathrm{w}} = \Rn^{-H/2}\Ry\Rn^{-1/2}.
\end{equation}
The \ac{evd} of \eqref{eq:prewhitening} is equal to
\begin{equation}
	\label{eq:evd}
	\Ry^{\mathrm{w}} = \mathbf{V}\bm{\Lambda}\mathbf{V}^H,
\end{equation}
where $\mathbf{V} \in \mathbb{C}^{M \times M}$ contains the eigenvectors and the diagonal matrix $\bm{\Lambda} \in \mathbb{R}^{M \times M}$ contains the eigenvalues. Based on the principal eigenvector $\mathbf{v}_{\rm max}$, the RTF vector can be estimated as \cite{Markovich2015}
\begin{equation}
	\label{eq:cw}
	\hcw = \frac{\Rn^{1/2}\mathbf{v}_\mathrm{max}}{\eo^T\Rn^{1/2}\mathbf{v}_\mathrm{max}}.
\end{equation}
\vspace{-0.3cm}
\subsection{Iterative methods}
Iterative CS-based and CW-based methods for RTF estimation have been proposed, which aim at reducing the computational complexity of the \ac{evd} by using the power iteration method (or von-Mises-Iteration) to calculate the principal eigenvector $\mathbf{v}_\mathrm{max}$.
Using the power iteration method on the pre-whitened microphone signal covariance matrix $\Ry^\mathrm{w}$ \cite{Krueger2011} or the speech covariance matrix $\Rx$ \cite{Varzandeh2017} yields the \ac{pm} estimators $\hpmcw$ and $\hpmcs$, respectively.
As mentioned in \cite{Krueger2011,Varzandeh2017}, one iteration per frame is typically sufficient for an online implementation.

\section{Incorporation of an external microphone}
\label{sec:Emic}
In addition to the local microphone array, we now assume the presence of an \textit{external} microphone that is spatially separated from the local microphones.
The \textit{extended} microphone signal vector, containing the microphone signals of the local microphone array and the external microphone signal, is given by
\begin{equation}
	\label{eq:extMicrophoneSignals}
	\by = \begin{bmatrix}
		\y\\
		Y_{\rm E}
	\end{bmatrix} \in \mathbb{C}^{M+1},
\end{equation}
where $\YE$ denotes the external microphone signal. The extended speech and noise vectors are defined similarly as $\bx = \left[ \x, \; \XE \right]^T$ and $\bn = \left[ \n, \; \NE \right]^T$, respectively.
Similarly to \eqref{eq:rtf}, the extended \ac{rtf} vector is given by
\begin{equation}
	\label{eq:extRtf}
	\bh = \begin{bmatrix}
		\h\\
		A_{\rm E} / A_{\rm 1}
	\end{bmatrix} \in \mathbb{C}^{M+1},
\end{equation}
where $A_\mathrm{E}$ denotes the \ac{atf} between the desired speech source and the external microphone.
Similarly to \eqref{eq:corrMatRx} and \eqref{eq:corrMatRn}, the extended speech covariance matrix and the extended noise covariance matrix are equal to
\begin{align}
	\label{eq:extSpeechMat}
	\bRx &= \mathcal{E}\{\bx\bx^H\} = \phixone \bh\bh^H \in \mathbb{C}^{(M+1) \times (M+1)},\\
	\bRn &= \mathcal{E}\{\bn\bn^H\} \in \mathbb{C}^{(M+1) \times (M+1)}.
	\label{eq:extNoiseMat}
\end{align}
Similarly to \eqref{eq:corrMatRy}, the extended microphone signal correlation matrix is equal to $\bRy = \bRx + \bRn$.
We assume that the distance between the external microphone and the local microphones is large enough such that the noise components in the local microphone signals are spatially uncorrelated with the noise component in the external microphone signal, i.e.,
\begin{equation}\label{eq:assum}
	\mathcal{E}\{\n\NE^*\} = \mathbf{0}_{M},
\end{equation}
where $\mathbf{0}_M$ is an $M$-element zero vector. Hence, the extended noise covariance matrix $\bRn$ in \eqref{eq:extNoiseMat} can be written as
\begin{equation}
	\label{eq:extSpatCoh}
	\bRn = \left[\begin{array}{c|c}
		\bRn & \mathbf{0}_{M}\\ \hline
		\mathbf{0}_{M}^T & \Phi_{\rm n_E}
	\end{array}\right],
\end{equation}
where $\Phi_{\rm n_E} = \mathcal{E}\{|N_{\rm E}|^2\}$ denotes the noise \ac{psd} in the external microphone signal.
For a diffuse, i.e., spherically isotropic, noise field, the \ac{sc} between the noise component in the external microphone signal and the noise component in a local microphone signal is equal to (neglecting head shadow effects)
\begin{equation}
	\gamma = {\rm sinc}\left(d \omega/c\right),
\end{equation}
with $d$ the distance between the external microphone and the local microphone, $\omega$ the angular frequency and $c$ the speed of sound.
Hence, the assumption in \eqref{eq:assum} already holds well even for relatively small distances (especially at high frequencies).\\
Based on the assumption in \eqref{eq:assum}, it can be easily shown that the covariance between the local microphone signals and the external microphone signal is equal to the covariance between the speech components in these microphone signals, i.e.,
\begin{equation}
\begin{split}
		\label{eq:corrLocExt}
		\mathcal{E}\{\y\YE^*\} &= \mathcal{E}\{\left(\x + \n\right)\left(\XE^* + \NE^*\right)\} = \mathcal{E}\{\x\XE^*\}.
	\end{split}
\end{equation}
Using the CS method described in Section \ref{sec:cs}, the extended RTF vector $\bh$ in \eqref{eq:extRtf} can be estimated as the last column of the extended speech covariance matrix $\bRx$, normalized by the first entry (corresponding to the reference microphone), i.e.,
\begin{equation}\label{eq:extCS}
  \bh = \frac{\bRx\beE}{\beo^T\bRx\beE}
\end{equation}
with the $(M+1)$-dimensional selection vectors $\beo = [1,0,\dots,0]^T$ and $\beE = [0,\dots,0,1]^T$.
Using \eqref{eq:extSpatCoh}, it can easily be shown that
\begin{align}
	&\bRy\beE = \bRx\beE + \bRn\beE = \bRx\beE + \Phi_{\rm n_E}\beE,\\
	&\beo^T\bRy\beE = \beo^T\bRx\beE + \beo^T\bRn\beE = \beo^T\bRx\beE,
\end{align}
such that, using \eqref{eq:extCS},
\begin{equation}\label{eq:estimH}
  \frac{\bRy\beE}{\beo^T\bRy\beE} = \frac{\bRx\beE + \Phi_{\rm n_E}\beE}{\beo^T\bRx\beE} = \bh + \frac{\Phi_{\rm n_E}}{\beo^T\bRx\beE}\beE.
\end{equation}
Hence, an unbiased estimation for the RTF vector $\h$ can be obtained as the first elements of the vector in \eqref{eq:estimH}, i.e.,
\begin{equation}
\boxed{
  \h_{\rm SC} = \left[\mathbf{I}_M, \; \mathbf{0}_{M}\right] \frac{\bRy\beE}{\beo^T\bRy\beE}
	}
\end{equation}
where $\mathbf{I}_M$ is the identity matrix of size $M$ and
which requires an estimate of the extended microphone covariance matrix $\bRy$ and no estimate of any noise covariance matrix.
The proposed estimator has a low computational complexity (similar to the CS estimator using only the local microphone signals), but obviously requires an external microphone signal to be transmitted to the local microphone array (synchronization aspects are outside the scope of this paper).
Assuming the availability of a \ac{vad} that outputs 1 if speech is present and 0 if speech is absent, the proposed RTF estimation algorithm is summarized in Algorithm \ref{algo:rtfEstimation}, where the extended microphone signal covariance matrix is recursively updated during speech-plus-noise frames. 

\begin{algorithm}
	\caption{Proposed RTF estimation algorithm}
	\label{algo:rtfEstimation}
	\begin{algorithmic}[1]
		\renewcommand{\algorithmicrequire}{\textbf{Input:}}
 		\renewcommand{\algorithmicensure}{\textbf{Parameter:}}
 		\REQUIRE $\by(l)$, $\bRy(l-1)$, $\mathrm{VAD}(l)$
 		\ENSURE smoothing factor $\alpha$\\
 		For each frequency bin:
 		\\ \textit{Estimation of the extended microphone signal covariance matrix:}
 		\IF{($\mathrm{VAD}(l)==1$)}
 		\STATE $\bRy(l) = \alpha \bRy(l-1) + (1-\alpha)\by(l)\by^H(l)$
 		\ELSE
 		\STATE $\bRy(l) = \bRy(l-1)$
 		\ENDIF
 		\\ \textit{Estimation of the RTF vector:}
 		\STATE $\h_{\rm SC}(l) = \left[\mathbf{I}_M, \; \mathbf{0}_{M}\right] \frac{\bRy(l)\beE}{\beo^T\bRy(l)\beE}$
 		\renewcommand{\algorithmicensure}{\textbf{Output:}}
 		\ENSURE  $\h_{\rm SC}(l)$
	\end{algorithmic}
\end{algorithm}
\vspace{-0.3cm}
\section{Experimental results}
In this section we compare the performance of the proposed RTF estimator (using the local and the external microphones) with all RTF estimations discussed in Section \ref{sec:esti} (using only the local microphone signals).
Section \ref{sec:setup} describes the experimental setup and the algorithmic parameters.
Section \ref{sec:rtfError} and \ref{sec:noiseReduction} evaluate the RTF estimation accuracy and the noise reduction performance when using the RTF estimates in an MVDR beamformer. 
\subsection{Experimental setup}
\label{sec:setup}
For the simulations we used the database of real-world recordings (sampling frequency $f_s=\SI{16}{kHz}$) described in \cite{Woods2015}.
The room dimensions were about $12.7 \times 10 \times 3.6$ \si{\m^3} with a reverberation time of about \SI{620}{ms}.
The local microphone array consisted of $M=4$ microphones mounted to the ears of a listener (two microphones per ear).
As reference microphone we chose the front microphone mounted to the left ear.
The external microphone was located on a table in front of the desired speaker with about $\SI{60}{cm}$ distance to the reference microphone.
The desired speaker was an English-speaking female talker who sat to the right of the listener at an angle of about \SI{45}{\degree}.
Both the listener and the desired speaker were seated at a circular table with a diameter of \SI{106}{cm}. 
In addition, 56 other talkers which were also seated at tables, generated a realistic babble noise. The noise field hence contained mainly diffuse but also directional components from temporally dominant interfering talkers.
Separate recordings of the babble noise and the desired speaker were used to mix them together at different input \acp{snr} $\{-10, \, -5,\, 0,\, 5,\, 10\} \: \si{dB}$.
The SNR in the external microphone signal was about \SI{13}{dB} higher than in the reference microphone signal (due to distance and head shadow effect).
We calculated all \acp{snr} using the \textit{intelligibility-weighted} SNR \cite{Greenberg1993}.
The total signal length was about \SI{23}{s}.\\
We used an \ac{stft} framework with a frame length $L_\mathrm{f} = 512$ samples and a frame-shift of $L_\mathrm{o} = L_\mathrm{f}/2$ samples and a square-root Hann window.
To estimate the covariance matrices $\bRy$ and $\Ry$ using speech-plus-noise frames and the noise covariance matrix $\Rn$ using noise-only frames we used a simple broadband energy-based \ac{vad} calculated from the speech component in the reference microphone signal.
To recursively estimate these covariance matrices, we used the time constants $\tau_\mathrm{y}$ and $\tau_\mathrm{n}$, respectively.
The corresponding smoothing factor (cf. Algorithm 1) is equal to $\alpha = \exp( - \frac{L_\mathrm{f} - L_\mathrm{o}}{f_\mathrm{s} \tau})$.
Please note that a smaller time constant corresponds to a smaller smoothing factor and hence to a faster adaptation to possible changes, but may also lead to less accurate estimates of the covariance matrices.
Especially in a scenario where the microphones or the desired speaker may change their position, a small time constant is desireable to be able to track changes fast enough.
Because the background noise can be assumed to be rather stationary we set the corresponding time constant $\tau_\mathrm{n} = \SI{500}{ms}$.
The time constant used to recursively estimate the covariance matrices $\bRy$ and $\Ry$ was chosen as $\tau_\mathrm{y} \in \{50, \, 100,\, 150,\, 200\} \: \si{ms}$.
All covariance matrix estimates were initialised using the corresponding long-term estimate.
\vspace{-1mm}
\subsection{RTF estimation accuracy}
\label{sec:rtfError}
\begin{figure}[t!]
	\includegraphics[width=1\linewidth]{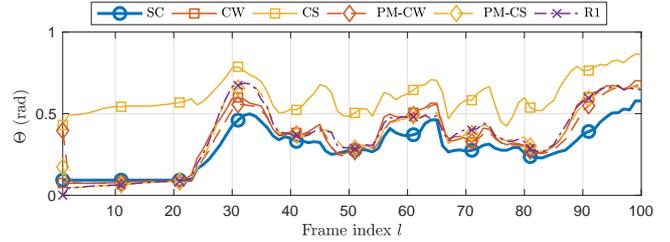}
	\center
	\vspace{-0.4cm}
	\caption{Hermitian angle $\Theta$ between the reference RTF vector $\href$ and the estimated RTF vectors (averaged over frequency) for an input SNR of 0 dB and $\tau_\mathrm{y}= \SI{50}{ms}$.}
	\label{fig:thetaFrames}
	\vspace{-0.3cm}
\end{figure}

As suggested in \cite{Varzandeh2017}, to evaluate the RTF estimation accuracy we used the Hermitian angle between a reference RTF vector $\href$ and the estimated RTF vector $\hat{\mathbf{h}}$, i.e.,
\begin{equation}
	\label{eq:HermitianAngle}
	\Theta(k,l) = \arccos \frac{|\href^H(k,l)\hh(k,l)|}{\|\href(k,l)\|_2\|\hh(k,l)\|_2}.
\end{equation}
The reference \ac{rtf} vector $\href$ was calculated as the principal eigenvector of the oracle speech covariance matrix $\Rx$ (estimated using all available speech frames), normalised by its first element (corresponding to the reference microphone).
\begin{figure}[t!]
	\includegraphics[width=0.991\linewidth]{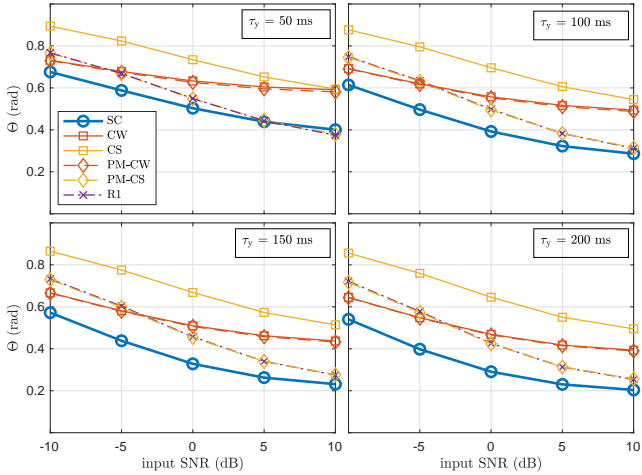}
	\center
	\vspace{-0.4cm}
	\caption{Hermitian angle $\Theta$ between the reference RTF vector $\href$ and the estimated RTF vectors (averaged over frequency and time) for different input SNRs and different time constants $\tau_\mathrm{y}$.}
	\label{fig:thetaEval}
	\vspace{-0.3cm}
\end{figure}
Figure \ref{fig:thetaEval} depicts the results (averaged over all frequencies and frames) for different time constants over different input \acp{snr}.
As expected, the performance of all estimators improves by increasing the input SNR and the time constant.
It can be observed that the proposed SC-based estimator generally outperforms the other estimators for all input \acp{snr} and time constants. The \ac{cs} method showed worse performance, in line with the literature \cite{Markovich2015}.
Only for a time constant of $\tau_\mathrm{y} = \SI{50}{ms}$ and a high input SNR of \SI{10}{dB} the R1 and PM-CS estimators slightly outperform the proposed estimator.
For an exemplary input SNR of $\SI{0}{dB}$ and a time constant of $\SI{50}{ms}$ 
Figure \ref{fig:thetaFrames} depicts the Hermitian angles (averaged over all frequencies) for the first 100 frames.
The proposed estimator starts to adapt after about 22 frames because this is the first frame where the speaker is active. All other estimators rely on estimates of both the noisy and the noise covariance matrices and hence adapt during noise-only and speech-plus-noise frames.
The R1 and CW estimators both seem to benefit from the long-term initializations in the first frames but perform worse than the proposed estimator afterwards.

\vspace{-0.3cm}
\subsection{Noise reduction}
\label{sec:noiseReduction}
\begin{figure}[t!]
	\includegraphics[width=0.991\linewidth]{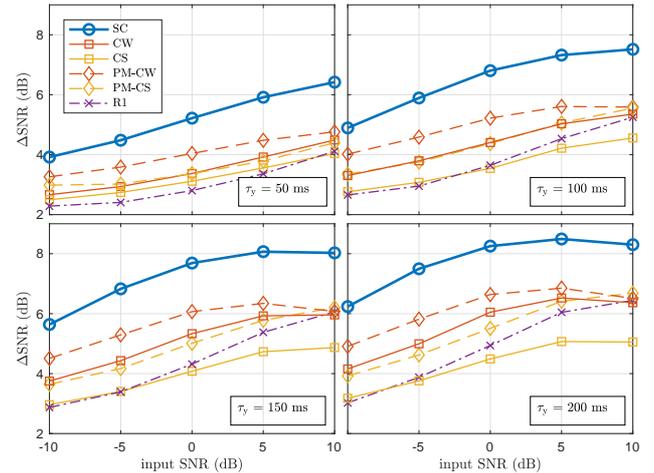}
	\center
	\vspace{-0.2cm}
	\caption{SNR improvement $\Delta$SNR of an MVDR beamformer steered by using the estimated RTF vectors for different time constants $\tau_\mathrm{y}$.}
	\vspace{-0.3cm}
	\label{fig:snrEval}
\end{figure}

We evaluated the noise reduction performance when using the estimated \acp{rtf} to steer an \ac{mvdr} beamformer, i.e., using $\hat{\h}(k,l)$ and the time-varying estimate of $\Rn(k,l)$ in \eqref{eq:mvdr}.
Please note, that for all estimators the MVDR beamformer is $M$-dimensional.
Figure \ref{fig:snrEval} depicts the \ac{snr} improvement ($\Delta$SNR) calculated by applying the beamformer to the desired speech and noise components separately.
As can be seen, the proposed \ac{sc} estimator clearly outperforms all other estimators for all input \acp{snr} and time constants.
\vspace{-0.3cm}
\section{Conclusion}
In this paper, we proposed an RTF estimation method exploiting low spatial coherence between the noise components in local microphone signals and an external microphone signal.
We derived a simple and computational efficient RTF estimator that yields an unbiased estimate of the RTF vector corresponding to the local microphone array.
Evaluation results in terms of the RTF estimation error and the noise reduction performance using real-world signals in an online implementation showed that the proposed estimator outperforms existing estimators using only the local microphone signals.
\bibliographystyle{IEEEbib}
\bibliography{journal1}

\begin{thebibliography}{10}

\bibitem{Doclo2015}
S.~Doclo, W.~Kellermann, S.~Makino, and S.E. Nordholm,
\newblock ``{Multichannel Signal Enhancement Algorithms for Assisted Listening
  Devices: Exploiting spatial diversity using multiple microphones},''
\newblock {\em IEEE Signal Processing Magazine}, vol. 32, no. 2, pp. 18--30,
  Mar. 2015.

\bibitem{Bertrand2009}
A.~Bertrand and M.~Moonen,
\newblock ``{Robust Distributed Noise Reduction in Hearing Aids with External
  Acoustic Sensor Nodes},''
\newblock {\em EURASIP Journal on Advances in Signal Processing}, vol. 2009,
  pp. 14 pages, Jan. 2009.

\bibitem{Markovich-Golan2015}
S.~Markovich-Golan, A.~Bertrand, M.~Moonen, and S.~Gannot,
\newblock ``Optimal distributed minimum-variance beamforming approaches for
  speech enhancement in wireless acoustic sensor networks,''
\newblock {\em Signal Processing}, vol. 107, pp. 4--20, Feb. 2015.

\bibitem{Szurley2016}
J.~Szurley, A.~Bertrand, B.~Van~Dijk, and M.~Moonen,
\newblock ``Binaural noise cue preservation in a binaural noise reduction
  system with a remote microphone signal,''
\newblock {\em IEEE/ACM Trans. on Audio, Speech and Language Processing}, vol.
  24, no. 5, pp. 952--966, May 2016.

\bibitem{Yee2017J}
D.~Yee, H.~Kamkar-Parsi, R.~Martin, and H.~Puder,
\newblock ``{A Noise Reduction Post-Filter for Binaurally-linked
  Single-Microphone Hearing Aids Utilizing a Nearby External Microphone},''
\newblock {\em IEEE/ACM Trans. on Audio Speech and Language Processing}, vol.
  26, no. 1, pp. 5--18, 2017.

\bibitem{Goessling2017}
N.~G\"o{\ss}ling, D.~Marquardt, and S.~Doclo,
\newblock ``Performance analysis of the extended binaural {MVDR} beamformer
  with partial noise estimation in a homogeneous noise field,''
\newblock in {\em Proc. Joint Workshop on Hands-free Speech Communication and
  Microphone Arrays (HSCMA)}, San Francisco, USA, Mar. 2017, pp. 1--5.

\bibitem{Ali2018}
R.~Ali, T.~Van~Watershoot, and M.~Moonen,
\newblock ``{Generalised sidelobe canceller for noise reduction in hearing
  devices using an external microphone},''
\newblock in {\em Proc. IEEE International Conference on Acoustics, Speech and
  Signal Processing (ICASSP)}, Calgary, Alberta, Kanada, Apr. 2018, pp.
  521--525.

\bibitem{Veen1988}
B.~D. Van~Veen and K.~M. Buckley,
\newblock ``Beamforming: a versatile approach to spatial filtering,''
\newblock {\em IEEE ASSP Magazine}, vol. 5, no. 2, pp. 4--24, Apr. 1988.

\bibitem{Gannot2001}
S.~Gannot, D.~Burshtein, and E.~Weinstein,
\newblock ``{Signal Enhancement Using Beamforming and Non-Stationarity with
  Applications to Speech},''
\newblock {\em IEEE Trans. on Signal Processing}, vol. 49, no. 8, pp.
  1614--1626, Aug. 2001.

\bibitem{Warsitz2007}
E.~Warsitz and R.~Haeb-Umbach,
\newblock ``{Blind acoustic beamforming based on generalized eigenvalue
  decomposition},''
\newblock {\em IEEE Trans. on Audio Speech and Language Processing}, vol. 15,
  no. 5, pp. 1529--1539, July 2007.

\bibitem{Markovich2009}
S.~Markovich, S.~Gannot, and I.~Cohen,
\newblock ``Multichannel eigenspace beamforming in a reverberant noisy
  environment with multiple interfering speech signals,''
\newblock {\em IEEE Trans. on Audio, Speech, and Language Processing}, vol. 17,
  no. 6, pp. 1071--1086, Aug. 2009.

\bibitem{Krueger2011}
A.~Krueger, E.~Warsitz, and R.~Haeb-Umbach,
\newblock ``{Speech enhancement with a {GSC}-like structure employing
  eigenvector-based transfer function ratios estimation},''
\newblock {\em IEEE Trans. on Audio Speech and Language Processing}, vol. 19,
  no. 1, pp. 206--219, Jan. 2011.

\bibitem{Marquardt2015a}
D.~Marquardt, E.~Hadad, S.~Gannot, and S.~Doclo,
\newblock ``{Theoretical Analysis of Linearly Constrained Multi-channel Wiener
  Filtering Algorithms for Combined Noise Reduction and Binaural Cue
  Preservation in Binaural Hearing Aids},''
\newblock {\em IEEE/ACM Trans. on Audio, Speech, and Language Processing}, vol.
  23, no. 12, pp. 2384--2397, Dec. 2015.

\bibitem{Hadad2016}
E.~Hadad, S.~Doclo, and S.~Gannot,
\newblock ``{The Binaural LCMV Beamformer and its Performance Analysis},''
\newblock {\em IEEE/ACM Trans. on Audio, Speech, and Language Proc.}, vol. 24,
  no. 3, pp. 543--558, 2016.

\bibitem{Hassani2016}
A.~Hassani, A.~Bertrand, and M.~Moonen,
\newblock ``{LCMV beamforming with subspace projection for multi-speaker speech
  enhancement},''
\newblock in {\em Proc. IEEE International Conference on Acoustics, Speech and
  Signal Processing (ICASSP)}, Shanghai, China, May 2016, pp. 91--95.

\bibitem{Taseska2016}
M.~Taseska and E.~A.~P. Habets,
\newblock ``{Spotforming: Spatial Filtering with Distributed Arrays for
  Position-Selective Sound Acquisition},''
\newblock {\em IEEE/ACM Trans. on Audio Speech and Language Processing}, vol.
  24, no. 7, pp. 1291--1304, 2016.

\bibitem{Koutrouvelis2018}
A.~Koutrouvelis, T.~W. Sherson, R.~Heusdens, and R.~C. Hendriks,
\newblock ``{A Low-Cost Robust Distributed Linearly Constrained Beamformer for
  Wireless Acoustic Sensor Networks With Arbitrary Topology},''
\newblock {\em IEEE/ACM Trans. on Audio Speech and Language Processing}, vol.
  26, no. 8, pp. 1434--1448, 2018.

\bibitem{Cohen2004}
I.~Cohen,
\newblock ``Relative transfer function identification using speech signals,''
\newblock {\em IEEE Trans. on Speech and Audio Processing}, vol. 12, no. 5, pp.
  451--459, Sep. 2004.

\bibitem{Markovich2015}
S.~Markovich-Golan and S.~Gannot,
\newblock ``{Performance analysis of the covariance subtraction method for
  relative transfer function estimation and comparison to the covariance
  whitening method},''
\newblock in {\em Proc. IEEE International Conference on Acoustics, Speech and
  Signal Processing (ICASSP)}, Brisbane, Australia, Apr. 2015, pp. 544--548.

\bibitem{Giri2016}
R.~Giri, D.~Rao, B., F.~Mustiere, and T.~Zhang,
\newblock ``{Dynamic relative impulse response estimation using structured
  sparse Bayesian learning},''
\newblock in {\em Proc. IEEE International Conference on Acoustics, Speech and
  Signal Processing (ICASSP)}, March 2016, pp. 514--518.

\bibitem{Varzandeh2017}
R.~Varzandeh, M.~Taseska, and E.~A.~P. Habets,
\newblock ``An iterative multichannel subspace-based covariance subtraction
  method for relative transfer function estimation,''
\newblock in {\em Proc. Joint Workshop on Hands-free Speech Communication and
  Microphone Arrays (HSCMA)}, San Francisco, USA, Mar. 2017, pp. 11--15.

\bibitem{Serizel2014}
R.~Serizel, M.~Moonen, B.~Van~Dijk, and J.~Wouters,
\newblock ``Low-rank approximation based multichannel {Wiener} filter
  algorithms for noise reduction with application in cochlear implants,''
\newblock {\em IEEE/ACM Trans. on Audio, Speech and Language Processing}, vol.
  22, no. 4, pp. 785--799, Apr. 2014.

\bibitem{Woods2015}
W.~S. Woods, E.~Hadad, I.~Merks, B.~Xu, S.~Gannot, and T.~Zhang,
\newblock ``A real-world recording database for ad hoc microphone arrays,''
\newblock in {\em Proc. IEEE Workshop on Applications of Signal Processing to
  Audio and Acoustics (WASPAA)}, New Paltz, New York, Oct 2015, pp. 1--5.

\bibitem{Greenberg1993}
J.~E. Greenberg, P.~M. Peterson, and P.~M. Zurek,
\newblock ``Intelligibility-weighted measures of speech-to-interference ratio
  and speech system performance,''
\newblock {\em Journal of the Acoustical Society of America}, vol. 94, no. 5,
  pp. 3009--3010, Nov. 1993.

\end{thebibliography}
\end{document}